\documentclass{ws-ijmpd}
\usepackage[super,compress]{cite}
\usepackage{graphicx}
\usepackage[numbers]{natbib}
\newcommand{\beq}{\begin{equation}}
\newcommand{\eeq}{\end{equation}}
\newcommand{\beqa}{\begin{eqnarray}}
\newcommand{\eeqa}{\end{eqnarray}}

\begin{document}
\markboth{Alexander F. Zakharov}{Tests of gravity theories  with Galactic Center observations}

%
\catchline{}{}{}{}{}
%

\title{Tests of gravity theories  with Galactic Center observations}

\author{Alexander F. Zakharov$^{1,2,3,4}$
}

\address{$^{1}$Institute of Theoretical and Experimental
Physics,
            B. Cheremushkinskaya 25,             117218 Moscow,  Russia\\
zakharov@itep.ru}

\address{$^{2}$Bogoliubov
Laboratory of Theoretical Physics, Joint Institute
  for Nuclear Research,
   141980 Dubna,
             Russia}

\address{$^{3}$National Research Nuclear University  MEPhI
(Moscow Engineering Physics Institute), 115409, Moscow,
 Russia}

\address{
$^{4}$North Carolina Central University, Durham, NC
27707,
 USA
 }

\maketitle

\begin{history}
\received{Day Month Year}
\revised{Day Month Year}
\end{history}

\begin{abstract}

An active stage of relativistic astrophysics started in 1963 since in this year, quasars were discovered, Kerr solution has been found and the first Texas Symposium on Relativistic Astrophysics was organized in Dallas. Five years later, in 1967--1968 pulsars were discovered and their model as rotating neutron stars has been proposed, meanwhile J. A. Wheeler claimed that Kerr and Schwarzschild vacuum solutions of Einstein equations provide
an efficient approach for astronomical objects with different masses. Wheeler suggested to call these objects black holes.
Neutron stars were observed in different spectral band of electromagnetic radiation. In addition, a neutrino signal  has been found for SN1987A.
Therefore, multi-messenger astronomy demonstrated its efficiency for decades even before observations of the first gravitational radiation sources.
However,
usually, one has only manifestations of black holes in a weak gravitational field limit and sometimes a model with a black hole could be substituted with an alternative approach which very often looks much less natural, however, it is necessary to find observational evidences to reject such an alternative model. At the moment only a few astronomical signatures for strong gravitational field are found, including a shape of relativistic iron $K_\alpha$ line, size and shape of shadows near black holes at the Galactic Centers and M87, trajectories of bright stars near the Galactic Center.
After two observational runs the LIGO-- Virgo collaboration provided a confirmation for an presence of mergers for ten binary black holes and one binary neutron star system where gravitational wave signals were found.
In addition, in last years a remarkable progress  has been reached in a development of observational facilities to investigate a gravitational potential,
for instance, a number of telescopes operating in  the Event Horizon Telescope network is increasing and accuracy of a shadow reconstruction near the Galactic Center is improving, meanwhile largest VLT, Keck telescopes with adaptive optics and especially, GRAVITY facilities observe bright IR stars at the Galactic Center with a perfecting accuracy. More options for precision observations of bright stars will be available with creating extremely large telescopes TMT and E-ELT.
    It is clear that the Galactic Center (Sgr $A^*$) is a specific objects for observations.
Our Solar system is located
at a distance around 8~kpc from the Galactic Center (GC). Earlier, theorists proposed a number of different models including exotic ones for GC such as
boson star, fermion ball, neutrino ball, a cluster of neutron stars. Later, some of these models are ruled out or essentially constrained with
consequent observations and theoretical considerations.
Currently, a supermassive black hole with  mass around $4 \times 10^6~M_\odot$ is the most natural model for GC.
Using results of observations for trajectories of bright stars  in paper \cite{Zakharov_JCAP_16} the authors got a graviton mass constraint which is comparable and consistent
with  constraints obtained recently by the LIGO-Virgo collaboration. Later, we consider opportunities to improve current graviton mass mass constraints with future observations of bright stars \cite{Zakharov_JCAP_18}. Similarly, from an analysis of bright star trajectories one could constrain a tidal charge which was  predicted by a gravity theory with an additional dimension \cite{Zakharov_EPJC_18}.

\end{abstract}

\ccode{PACS numbers: 04.20.-q, 04.30.Tv, 04.70.-s, 04.70.Bw,
04.80.Cc, 98.35.Jk}


\section{A history of black hole concept}

At the moment there is no consensus concerning a starting point for black hole concept, some people say that it was introduced at the end of XVIII century in studies of the Reverend John Michel and Pierre  Simon Laplace \cite{Michell_84,Laplace_96,Laplace_99}, some people say that it was started since the Schwarzschild paper \cite{Schwarzschild_16}\footnote{An interesting discussions of historical aspects of the issue are presented in \cite{Eisenstaedt_82,Montgomery_09}.}
It is well-known that one of the first solutions of general relativity was spherically symmetric vacuum solution found by Schwarzschild  \cite{Schwarzschild_16}, however, properties of the solution are rather unusual and it was a natural question about an opportunity to use such a solution for real physical objects. In 1939 A. Einstein investigated the issue and arrived at a negative conclusion that the Schwarzschild singularities (Einstein called the event horizon in such a way)
do not exist in physical reality \cite{Einstein_39}.
Later, Einstein's assistant Peter Bergmann supported the Einstein opinion about a unreality of Schwarzschild solution in one of the first  books on general relativity \cite{Bergmann_42}. As it was found in  thirties of the XX century
the final fate of star depends on its mass for white dwarfs and neutron stars. Soon after the discovery of Fermi -- Dirac statistics
in 1926 \cite{Fermi_26,Dirac_26}, Fowler supposed that degenerate electron gas pressure is counterbalancing an action of gravity \cite{Fowler_26}, while soon after that Soviet theorist Ya. Frenkel evaluated typical mass range for white dwarfs in \cite{Frenkel_28}. A detailed analysis of the Frenkel's paper and its consequences for an existence of mass limits of white dwarfs are presented in \cite{Yakovlev_94}. British physicist E. Stoner was the first researcher who found in the simple model of mass density distribution that white dwarfs should have upper mass limit \cite{Stoner_30} and later this statement was generalized
by  S. Chandrasekhar and L.D. Landau \cite{Chandrasekhar_31,Landau_32,Chandrasekhar_34}.
In paper \cite{Yakovlev_13}, the authors considered Landau's paper \cite{Landau_32} and related historical content and the authors concluded that
  Landau contributed a lot in a development of the neutron star concept in spite of the fact that words "neutrons" or "neutron stars"  were not used
  in \cite{Landau_32}, moreover, Landau noted that the conclusion about an upper mass limit for white dwarfs looks so extraordinary that there are regions in stars where laws of quantum mechanics and quantum (Fermi -- Dirac) statistics should be violated.
    Soon after the neutron discovery Baade and Zwicky outlined a scenario for  formations of neutron stars could be observed as supernova explosions \cite{Baade_34}. Later G. Gamow  concluded that similarly to white dwarfs neutron stars should have an upper limit for their masses \cite{Gamow_37}. In his consideration Gamow used Newtonian approximation for a gravitational field of neutron stars and the analysis has a numerical mistake which was corrected in  \cite{Ludwig_14}.
   After a few years,  J. R. Oppenheimer and G. M. Volkoff generalized the Gamow's result considering gravity in a general relativity approach \cite{Oppenheimer_V_39}, meanwhile J. R.  Oppenheimer and H. Snyder analyzed a spherically symmetric collapse of a dust ball which is forming  a Schwarzschild vacuum solution \cite{Oppenheimer_39}, however at this time people thought that such theoretical model is too simple and artificial  to be correct because in their solution dust ball formed space-time singularity in a finite interval of a proper time.

In 1963 the first star-like object with large redshift  has been discovered \cite{Schmidt_63}. This object was observed earlier with Cambridge radio survey and it was called 3 C273. It  was observed in a survey  with Parkes Radio Telescope by Lunar occultation method \cite{Hazard_63}. M. Schmidt      evaluated the redshift of this object and it was rather high $z \approx 0.158$, while J. B. Oke  estimated a flux \cite{Oke_63}. It turned out that its luminosity in around 100 times higher than our Galaxy and immediately
astronomers and astrophysicists started to discuss possible physical processes to provide so huge energy release.
Analyzing old photo plates where was 3 C273 image, Soviet astronomers  and A. S. Sharov and Yu. N. Efremov found variabilities with time scale  $T_{var}$ around a few days \cite{Sharov_63}. It means that a size 3 C273 is less than $c T_{var}$ (where $c$ is a speed of light) or less than the Solar system size. It was a very challenging task to propose a theoretical model for a huge energy release from a very small space region.


 In December 1963 the first Texas Symposium on Relativistic Astrophysics has been organized in Dallas, quasars were discovered a few months before it and Roy Patrick Kerr found his solution which describes  rotating black holes. In 1964 the term "quasar" was introduced by  Hong-Yee Chiu as a short version of "quasi-stellar radio source" in his detailed review about the First Texas Symposium on Relativistic Astrophysics \cite{Chiu_64}.
 Astronomers and astrophysicists analyzed different possibilities to explain typical quasar features and excluding very exotic points of view there is an opinion that this
high luminosity is connected with a conversion of binding energy of accreting matter into radiation or with an opportunity to transform rotational energy of black hole
in radiation. In both cases a rotating black hole solution found by R. P. Kerr in 1963 is the important component to provide a huge energy release from a very small space region. A creation of quasar energy release is a bright example of efficiency of interaction scientists working in astrophysics and
general relativity.  Kip Thorne reminded \cite{Thorne_94}
that initially astronomers and astrophysicists were rather skeptical in respect to exact solutions of Einstein equations, but fortunately,  Achilles Papapetrou paid a special attention to Kerr solution. In his concluding remarks T. Gold expressed his hope that a fruitful interaction between
astrophysicists and relativists will continue.

 An introduction of term "black hole"  is associated  with J. A. Wheeler, but such a concept has been used earlier at the first Texas Symposium on Relativistic astrophysics in December 1963 as it was explained by A. Rosenfield in an article for "Life" magazine published on January 24, where he discussed the Hoyle -- Fowler
model, namely "as Hoyle theorized, gravitational collapse gravitational collapse would be "catastrophe implosion" on a cosmic scale", but criticizing the Hoyle's model and demonstrating its weaknesses, Rosenfield wrote "... instead of intensely radiating object sending out lavish quantities of light and radio energy, gravitational collapse would result it in invisible "Black hole" in the universe" \cite{Rosenfield_64}.

Before a publication of the "Life" issue with the Rosenfeld's article, on January 18, 1964 Ann Ewing reported about the meeting of the American
Association for the Advancement of Science
in Cleveland in January 1964 with a short article  "Space may be peppered with "black
holes""\footnote{These grains of pepper for nearby supermassive black holes in the Galactic Center and in the center of M87 could be observed as shadows (small dark spots in the sky) with the Event Horizon Telescope.} as a title \cite{Ewing_64}, therefore, clearly that in December 1963 at the First Texas Symposium and in January 1964 at the AAAS meeting astronomers and relativists discussed black holes in sky, however it is still unknown exactly who were these persons, see, also discussions in book \cite{Bartusiak_15} and in article "50 years later, it's hard to say who named black holes" by
T. Siegfried in internet journal "Science News" where the author wrote "but it didn't catch on until Wheeler began using it a few years later.
"Perhaps Wheeler still gets credit," Bartusiak said. "He never said he originated the term…. What was important is that he had the authority to give the scientific community permission to use the term 'black hole.'" \footnote{https://www.sciencenews.org/blog/context/50-years-later-it\%E2\%80\%99s-hard-say-who-named-black-holes.}
Bartusiak had conversations with Hong-Yee Chiu who was a member of the Institute of Advanced Studies on Princeton from 1959 until 1961 and based on his reminding,  it was noted that
R. Dicke discussed a collapse of massive stars with a formation of something similar to "black hole of Calcutta" \cite{Bartusiak_15}. A description of this historical event could find  see for instance, in records by
 John Zephaniah Holwell \cite{Holwell_74} where the author reminded the historical case when
after the fall of Fort Williams    Siraj ud-Daulah, the Nawab of Bengal,  ordered to place many British prisoners of war at a very small room in June 1756 (see also a description of the vent in a popular book \cite{Narlikar_82}). Really, it is a nice illustration of a huge concentration of  mass in a very limited space region.

Therefore, according to Bartusiak's opinion probably Dicke introduced the term for the scientific community \cite{Bartusiak_15}.
J.A. Wheeler was reminding circumstances which were stimuli to introduce the black hole concept \cite{Wheeler_00}. When information about the discovery reached US scientific community, Wheeler started to think about possible theoretical model  for pulsars discovered in UK in 1967 (the corresponding paper was published with a delay only in 1968 \cite{Hewish_68}
since initially these observational data were classified) and Wheeler understood that a pulsar model had to include "gravitationally completely collapsed object". In the autumn 1967, he was invited to attend a conference where possible interpretations of pulsar phenomenon were discussed and
Wheeler expressed his concern about unsuccessful searches for suitable terminology for "completely collapsed object" and somebody in audience suggested "How about a black hole". According to Wheeler's opinion it was extremely useful and fruitful suggestion and Wheeler used this term in his
Sigma Xi–Phi Beta Kappa lecture in the West Ballroom of the New York Hilton a few weeks later, on December 29, 1967 and it was included in his corresponding paper
in the spring of 1968 \cite{Wheeler_68}. Now it is known that a pulsar phenomenon is powered by a neutron star as it was discussed before the pulsar discovery by F. Pacini \cite{Pacini_67} and soon after the discovery by T. Gold  \cite{Gold_68}.\footnote{The standard model in particle physics has been proposed in 1967, see an overview of foundations of the model and its development in \cite{Weinberg_18}, therefore, 1967--1968 were very for physics.} F. Pacini and T. Gold worked in Cornell University and clearly it was not a pure incident that two principal authors of the pulsar model were from the same institution.
From this piece of history of black hole concept and pulsar phenomenon, one could conclude if a great mind (J.A. Wheeler in our case) would started to  think about an important challenging problem, he (she) often could find a scientific brilliant or in other words something which is very important even in the case of a failure to reach an initial goal.

In his paper \cite{Wheeler_68}, where  Wheeler used the term "black hole" in the written form at the first time, he  considered the Crab Nebula which is associated with a  supernovae remnant SN 1054 as a clear confirmation of Baade -- Zwicky scenario for neutron star formation where it was suggested that neutrons stars could be formed in supernova explosions \cite{Baade_34}. Really the object has been observed by Chinese astronomers for 21 months since July 4, 1054 (according to the our current chronology). Clearly, that one thousand years ago Chinese astronomers and astrologists used their own chronology and their own names for constellations. Dutchman sinologist J. J. L. Duyvendak identified the Guest Star described in Chinese records with Crab Nebula  \cite{Duyvendak_42},  see also consequent astronomical discussions given by  N. U. Mayall and J. H. Oort \cite{Mayall_42}. At this time when these two papers were prepared Duyvendak and Oort worked in   University of Leiden (Oort worked in the Observatory of the University), while Mayall worked in the Lick Observatory (California, USA) and had intensive communications with  Baade.
Soon after a publication of these two papers  \cite{Duyvendak_42,Mayall_42} W. Baade and R. Minkowski studied properties of Crab Nebula and concluded that Chinese Nova 1054 A. D. was a supernova of type I \cite{Baade_42,Minkowski_42}.
Later, the Crab nebula was identified with a radio source in 1963 and as a X-ray source in 1964 and as a  pulsar in 1968 (see also discussion of the source in \cite{Shklovskii_68}.
There are a couple of conclusions from Wheeler's considerations in  \cite{Wheeler_68}. First, sometimes,
people could recognize a significance and justify an  importance of some activity (observations of ancient Chinese astronomers in our case) for
fundamental science only after centuries (or even Millennium) and at this period one could think that the activity was useless for fundamental study but it was not. Second, a scientific knowledge is a result  of activity  of  enthusiastic people working in different areas, therefore, multi-disciplinary investigations are often very fruitful. In addition, one could note that rather routine observations almost 1000 years ago, their careful storage,  translations and interpretations of old historical records give crucial contribution in studies of nuclear matter which are still very important and interesting for a scientific community. Third, a scientific process is very fragile and it may be terminated due to many different reasons, for instance, old historical records may be destroyed, forgotten or ignored and in this case a confirmation of the Baade-- Zwicky scenario for NS formation will be incomplete. In a more general case action or inaction could lead to a complete solution (or failure) of a puzzle for a phenomenon of Nature or to a case when pieces of the puzzle will be blank, fuzzy or be in disorder.

Properties of nuclear matter, quark -- gluon plasma are still very important for theoretical studies and experimental science and
 a rapid development of Mega Science facilities for studies of nuclear matter such FAIR (Facility for Antiproton and Ion Research) (Darmstadt, Germany) and Nuclotron Based Facility (NICA,  Dubna, Russia) is  this claim. Therefore, as we discussed, even ancient  observations of stars  have an important value for fundamental physics and nuclear physics, in particular, but a justification of astronomical observations for a fundamental science could come many years later.

It is clear  that a suitable  terminology is extremely important and  the term "black hole" is used in different context. For instance,
Google found more than $10^9$ documents where people used this term, therefore it is very popular not only in science.
 However,  Wheeler's suggestion to use the black hole term consists of two important issues, first, in spite of infinite time of gravitating object for complete gravitational collapse, a distant observer could use
a limiting metric which is  result of the collapse, since after a finite time interval differences between a dynamical metric of collapsing matter and
a static or stationary metric which describes a result of gravitational collapse are very small, therefore, one could use a simpler static (stationary) model. Second, Wheeler believed that black holes have to exist as results of stellar collapses and as engines in AGNs and quasars.  At the end of this section I would like to point out an interesting paper \cite{Herdeiro_18} where historical aspects of a black hole concept development were also discussed.  Recently a relativistic astrophysics development before 1940 was discussed in \cite{Bonolis_17} in a more wider context.

\section{Astrophysical  black holes}

It is clear that classical black holes should not emit particles and photons, but in 1974 S. Hawking found that black holes should emit particles due to an existence of quantum tunneling \cite{Hawking_74}
(many years before the Hawking's paper publication famous Soviet theorist  V. N. Gribov  argued in his conversations with Ya. B. Zeldovich that black holes have to radiate \cite{Anselm_98,Khriplovich_05}). The Hawking radiation could be important for black holes with masses significantly smaller than stellar ones. Such black holes may be formed at the early stage of Universe evolution and these black holes are called primordial ones.
     The Hawking radiation for black holes with stellar masses and heavier is negligible in comparison with other astrophysical processes. An opportunity to discover massive black holes with observations of electromagnetic radiation from accreting matter has been discussed by
E. Salpeter \cite{Salpeter_64} and Ya. B. Zeldovich \cite{Zeldovich_64}. Evaluating the energy for inner most stable circular orbit in the Schwarzschild metric one could conclude that an energy release could be a few percent of accreting matter mass (or which is comparable with nuclear fusion energy release), while for extreme Kerr metric such an energy release could be almost 50\% of mass or comparable with energy release in annihilation.

\subsection{Black holes with stellar masses}
Almost independently with the first X-ray observations of astronomical sources theoretical models of disk accretion have been developed in \cite{Pringle_72,Shakura_72,Shakura_73,Novikov_73}\footnote{Recently one of the pioneers of these theoretical studies  N. I. Shakura published interesting historical remindings of an initial period of
X-ray astronomy \cite{Shakura_14,Shakura_18}.}  and in these papers it was predicted that black holes and neutron stars with stellar masses could be found in X-ray stellar binary systems. Several BH systems
 have been found  in X-ray binaries with masses ($M_{BH} = 4-16 M_\odot$) and around one hundred neutron stars (NSs) as X-ray binaries, see for instance, papers
 \cite{Corrall_16,Cherepashchuk_16,Cherepashchuk_17}. It we take a look at a mass distribution of compact objects which are black holes and neutron stars in binary stellar systems, we could see that there is a clear mass gap between NSs and BHs with stellar masses and at the first glance, such a property does not look natural.  Sometimes even alternative theories of gravity have been used to explain such a phenomenon \cite{Sokolov_15}, however a fine tuning procedure for parameters of population synthesis gives an opportunity to solve such a puzzle \cite{Belzynski_12,Wiktorowicz_13}.

\subsection{Supermassive black holes in galactic centers}
Supermassive black holes with masses in range $(10^6 - 10^{10})M_\odot$ exist in quasars, active galactic nuclei and in centers of spiral galaxies
and and in this case they  could evolve in interactions with their galaxies  \cite{Kormendy_13}.  In active galactic nuclei black hole masses are evaluated with the reverberation method or a spectroscopy of absorption and emission lines.
Namely,  variations in the strengths of the central source in a quasar will
generate variations in the strengths and profiles of the emission lines. These "reverberations" (or "echos") in the
emission lines will delay in respect to continuum variations.
The reverberation method was suggested in \cite{Blandford_82}, however,  similar ideas were proposed earlier in \cite{Cherepashchuk_73}, see also  papers \cite{Gaskell_86,Cherepashchuk_16,Cherepashchuk_17} for  more recent reviews.

\subsection{Primordial black holes}
As it was proposed many years ago in \cite{Zeldovich_67,Hawking_71}, at the early stage of the Universe evolution black holes could be formed with rather wide mass spectrum  and in particular, their masses could be rather small and Hawking radiation should be significant for such objects \cite{Carr_74}. These black holes are called primordial or PBHs \cite{Polnarev_85}. Currently, PBHs are not discovered yet, however, they are very attractive as preferable models for many astrophysical puzzles such as a formation of supermassive black holes with high redshifts and discoveries of relatively heavy binary black holes with LIGO -- Virgo gravitational interferometers
\cite{Dolgov_18}.\footnote{Recently with data analysis of two observational runs the LIGO--Virgo collaboration reported about discoveries of ten gravitational wave signals from binary black hole mergers and one signal from binary neutron star merger, see \cite{Abbott_18}.} If such binary black holes are results of stellar evolution then their progenitors should have very strong stellar winds and binary black holes are placed not in vacuum but in matter and at the moment of a final merger of black holes an gravitational wave energy around $3 M_\odot c^2$ (as for the first GW event GW150914) was released as strong gravitational wave in a fraction of second, then  it is naturally to expect to detect an electromagnetic counterpart. It was proposed PBHs with intermediate masses as objects forming dark matter \cite{Chapline_16}. Astrophysical applications of PBHs with different masses are considered in a number of reviews \cite{Khlopov_10,Belotsky_14,Carr_16} (see also references therein).
Observational consequences of PBH clusters were considered in \cite{Belotsky_18}.

\section{Observations of bright stars near the Galactic Center}


The supermassive black hole  in our Galactic Center is the closest one, therefore, this object is very attractive and astronomers observe the Galactic Center in different spectral band including $\gamma$, X-ray, IR, optical, radio and mm-band \cite{Eckart_05,Dokuchaev_15a,Dokuchaev_15b,Zakharov_IJMPD_18,Zakharov_MIFI_17,Dokuchaev_18a,Dokuchaev_18b}.
Moreover, such an object is a natural laboratory to test general relativity and its alternatives in a strong gravitational field limit
\cite{Zakharov_05,Zakharov_07,Nucita_07,Borka_12,Zakharov_14}.
The GRAVITY collaboration observed S2 star pericenter passage in May 2018  reported about the discovery
of general relativity effects for this star \cite{Gravity_18}.
Currently there are two groups observing bright stars near the Galactic Center with largest telescopes equipped with adaptive optics facilities.
One (US) group uses the twin Keck telescopes with 10 m diameters at Hawaii, another ESO-MPE group uses four VLT telescopes with 8 m diameters
and now with GRAVITY facilities.
Roughly speaking, results of these two groups are consistent and complimentary.
Results of observations showed that in the first approximation stars are moving  along elliptical orbits
and therefore, one could conclude that motions of these stars are fitting rather well with a potential of point like mass around $M_{SBH}=4\times 10^6~M_\odot$ in the framework of Newtonian gravity law. One of the most interesting probes for a gravitational potential at the Galactic Center is S2 star. It has eccentricity $e = 0.88$,  period $T=16$~yr and an expected visible relativistic precession of its orbit is around $\Delta s \approx 0.83$~mas \cite{Gillessen_17,Chu_18} in assumption that extended mass distributions such as a stellar cluster or dark matter near the Galactic Center do not have a significant impact on relativistic precession of its orbit \cite{Zakharov_07,Nucita_07}. The Keck uncertainties in the S2 star orbit reconstruction are around $\sigma_{Keck}\approx 0.16$~mas \cite{Hees_PRL_17}, while for Thirty Meter Telescope(TMT) which will be constructed within a several years   uncertainties are $\sigma_{TMT}\approx 0.015$~mas.

\section{GRAVITY: first discoveries}

\subsection{Gravitational redshift for S2 star}

Both teams (Keck and VLT ones) continuously improve a precision for S2 star orbit reconstruction for both teams, for example, for MPE--ESO team operating with VLTs, since in 1990s a precision of SHARP facilities was around 4~mas, in 2000s NACO had a precision around 0.5~mas, but in 2018 GRAVITY reached a precision around $30~\mu as$ \cite{Gravity_18}.
In May 2018 the GRAVITY collaboration evaluated a gravitational redshift in the orbit of S2 star near its pericenter passage. In addition,  the GRAVITY collaboration estimated  relativistic precession of its orbit and showed that observational data are much better fitted with GR model in the first PN approach than in comparison with Newtonian one.
The authors introduced a parameter $f$ and $f=0$ corresponds to the Newtonian approach for gravitational redshift while $f=1$ corresponds to the first
post-Newtonian approximation of general relativity.
Therefore, general relativity successfully passed an additional test. It means that almost after 100 years after the confirmation  of the GR prediction about a deflection of light during Solar eclipse in 1919 \cite{Dyson_20} when astronomers tested three different options, namely, a) a deflection of light describes with the GR law; b) a deflection of light describes with the Newtonian law, and c) there is no a deflection of light in gravitational field. Now astronomers checked GR prediction in much stronger gravitational field at high distances from our Solar system and Einstein's theory of relativity successfully passed one important test more  \cite{Gravity_18}.
A theoretical approach for gravitational redshift evaluation if sources are moving in binary system was developed in \cite{Kopeikin_99,Alexander_05,Zucker_06}.
In May 2018 S2 star passed its pericenter and the GRAVITY collaboration established  that relativistic corrections have to be taken at the period near this passage. Near its pericenter S2 star had a total space velocity $V_{\rm peri} \approx 7650$ km/s or $\beta_{\rm peri} = V_{\rm peri}/c=2.55 \times 10^{-2}$ \cite{Gravity_18}.\footnote{Recently, it was reported precise measurements of gravitational redshifts with Galileo satellites \cite{Hermann_18}, an opportunity to test GR predictions with accuracy around $10^{-5}$ for gravitational redshifts using Radioastron observational data discussed in \cite{Litvinov_18}.}
Computing PPN(1) correction for a total gravitational redshift  following \cite{Kopeikin_99,Alexander_05,Zucker_06,Gravity_18}
\begin{equation}
 z_{\rm GR}=\frac{\Delta \lambda}{\lambda}=B_0+B_{0.5}\beta+B_1\beta^2+\mathcal{O}(\beta^3),
\label{redshift_1}
\end{equation}
where $B_1=B_{1,tD}+B_{1,grav}$, $B_{1,tD}=B_{1,grav}=0.5$, $B_{1,tD}$ is the special relativistic transverse Doppler effect,
$B_{1,grav}$ is the general relativistic gravitational redshift , $B_{0.5}=\cos \theta$, where $\theta$ is the angle between the velocity vector and line of sight \cite{Alexander_05}, while the total redshift $B_0$ which is independent on a star velocity $\beta$
\begin{equation}
  B_0 = z_\odot + z_{\rm gal}+z_{\rm star} +\frac{1}{2} \Upsilon_0,
  \label{redshift_2}
\end{equation}
therefore, the redshift $B_0$ consists of four terms, namely, $z_\odot$ is a redshift due a total motion of the Sun and the Earth in respect to Galactic Center and blue shift due to potential of the Sun and the Earth, $z_{\rm gal}$ is a redshift due to Galaxy potential, $z_{\rm star}$ is redshift due to the star's potential, the term $\dfrac{1}{2}\Upsilon_0 =\dfrac{GM}{2a}$ is due to the location of star in the SMBH potential \cite{Alexander_05}.
The GRAVITY collaboration estimated the total redshift from spectroscopical observations and concluded that it corresponds to $z \approx \dfrac{200~{km/s}}{c}$ \cite{Gravity_18}.
One could represent the total redshift obtained from spectroscopical observations in the form \cite{Gravity_18}
(as it was noted earlier $f$ reflects a weight of post-Newtonian term) or more precisely,
\begin{equation}
 z_{\rm tot}=z_K+f(z_{GR}-z_K),
\label{redshift_3}
\end{equation}
where $z_K=B_0+B_{0.5}\beta$ is the Keplerian redshift, $f=0$ corresponds to Keplerian (Newtonian) approach, while $f=1$ corresponds to PPN(1) approach.
The GRAVITY collaboration found that $f=0.90\pm 0.09|_{\rm stat}\pm 0.15 |_{\rm sys}$ and the authors also concluded that S2 data are inconsistent
with a pure Newtonian dynamics. Since $f$-value is slightly less than its expected value estimated with pure PPN(1) fit, perhaps an extended mass distribution of stellar cluster or dark matter should be taken into account in this model and in this case future observations of relativistic redshifts
and astrometric monitoring the bright stars
 will
help to evaluate parameters of an extended mass distribution and estimate a contribution of time dependent terms in a gravitational potential near the supermassive black hole.
Comparing fits for Schwarzschild and Newtonian precessions for a point like mass potential and observational data for S2 star, the GRAVITY collaboration evaluated $f$-value from observational data  (without any precession) and they concluded that the $f$-value is much closer to GR quantity $f=1$ or more precisely, the GRAVITY collaboration found $f=0.94 \pm 0.09$.

\subsection{Observations of motions of hot spots near ISCO}
The GRAVITY collaboration observed trajectories of two bright flares  on July 22 and July 28, 2018, as well as an orbit of fainter flare
on May 27, 2018 \cite{GRAVITY_18_spots}. The authors claimed that the position centroids exhibited clockwise looped motion on the sky, on scales of typically $150~\mu as$ over a time interval around a few tens of minutes, corresponding to about 30\% the speed of light. Meanwhile, the flares exhibited continuous rotation of the polarization angle, with about the same $45( \pm 15)$~min period as that of the centroid motions. Typical radius of spot orbits are around $7~M_{SBH}$ (in mass units), while  the innermost stable circular orbit (ISCO) radius is $6~M_{SBH}$ for a Schwarzschild black hole. Since there is a strong demand to explain these observational data
quantitatively, it is expected that a detailed theoretical model describing polarization variations of hot spots would be created shortly.

\subsection{Spacially resolved rotation of broad line region for 3C273}
Recently GRAVITY collaboration observed the first quasar 3C273 and found a spatial offset (with a spatial
resolution of $10^{-5}$ arcseconds, or about 0.03 parsecs for a distance
of 550 million parsecs) between the red and blue photo-centres of
the broad Paschen-$\alpha$ line of the quasar 3C 273 perpendicular to the
direction of its radio jet \cite{Gravity_Nature_18}.  The authors  fitted with a
broad-line-region model of a thick disk of gravitationally bound
material orbiting a black hole of $3 \times 10^{8}$ solar masses and they concluded
that disk radius is around 150 light days and earlier a disk radius in the range 100--400 light days was
found previously using reverberation mapping, therefore, new estimates are consistent with previous  ones.

\section{Constraints on alternative theories of gravity with observations of bright stars}

\subsection{Massive graviton  constraints}

A few years ago F. Dyson considered opportunities to detect a graviton \cite{Dyson_13} and his answer was rather pessimistic
since in the future it will be very hard to detect graviton in  physical experiments (or observations), independently on
 a version of gravity theory where graviton could be massive or massless.
However,
if graviton is massive there a number of different ways to constrain a graviton mass from astronomical
observations could be used \cite{Goldhaber_10,deRham_17}.
Initially theories of massive gravity have pathologies including discontinuities and a presence of ghosts, however, recently
theorists found ways to create ghost-free theories of massive gravity \cite{deRham_17}.
In 2016 the LIGO--Virgo collaboration reported about the first detection of
gravitational waves from a merger of a binary black hole system (the event was detected
on September 14, 2015 and it is called GW150914) \cite{Abbott_16}.
Moreover, the LIGO--Virgo collaboration considered a theory of massive gravity as an alternative for
conventional general relativity and found a constraint for  the graviton Compton wavelength
$\lambda_g > 10^{13}$~km which could be interpreted as a constraint
for a graviton mass $m_g < 1.2 \times 10^{-22}$~eV \cite{Abbott_16}.
 Later, the LIGO--Virgo collaboration  LIGO   reported about the discovery of the third GW event from merging the BHs with 31 and 19 solar masses at redshift $z=0.19$ observed on January 4, 2017,  named GW170104 and the authors significantly improved an upper graviton mass constraint
$m_g < 7.7 x 10^{-23}$~eV \cite{Abbott_17}.
In August 2017 a gravitational wave signal has been detected from binary neutron star merger
(GW170817). This source was observed by space spacecrafts Fermi and INTEGRAL and many ground based facilities, including global robotic network Master \cite{Lipunov_17}.
The LIGO -- Virgo collaboration together with its partner groups observing the same astronomical sources with different facilities found that
constraints on
speed of gravitational waves from binary neutron star merger (GW170817) are
$-3\times 10^{-15} < (v_{g}-c)/c < 7 \times 10^{-16}$   \cite{Abbott_17c}. Since graviton
energy is $E=hf$, therefore, assuming a typical LIGO frequency range
$f \in (10,100)$, from the dispersion relation one could obtain a
graviton mass estimate $m_g < 3 \times (10^{-21}-10^{-20})$~eV which is
a weaker estimates than previous ones obtained from binary
black hole signals detected by the LIGO team
\cite{Zakharov_IHEP_2017}.
In the case of massive graviton, one could use Yukawa gravitational potential in a
form $\propto r^{-1}\exp(-r/\lambda_g)$,
and in this case a lower
bound for Compton wavelength $\lambda_g$ of the graviton is connected with an
upper bound of its mass
\begin{equation}
m_{g(upper)}=h\,c/\lambda_g.~
\label{mass_vs_lambda}
\end{equation}
Some time ago, in paper \cite{Borka_13} we obtained constraints on Yukawa gravity from observational data for the S2 star orbit.
Later, we found constraints on graviton mass  $m_g < 2.9 \times 10^{-21}$~eV from available observational data \cite{Zakharov_JCAP_16} (see also \cite{Zakharov_Quarks_16,ZJBBJ_MIFI_17,ZJBBJ_Baldin_17} for more details). In these considerations we used available and relatively old
observational data to constrain a graviton mass. Later, Keck group followed our ideas to improve our estimates with new observational data $m_g < 1.6 \times 10^{-21}$~eV \cite{Hees_PRL_17}.
In paper \cite{Zakharov_JCAP_18} we considered  perspectives to improve current graviton mass estimates found with future observational data for S2 and other bright stars observed with VLT and Keck telescopes, in particular, we  evaluated orbital precession for Yukawa potential and  obtained
an upper limit for a graviton mass assuming that GR prediction about orbital precession will be confirmed with future observations.
In this case as it was shown in \cite{Zakharov_JCAP_18} the longest Compton wavelength could be expressed as
\begin{equation}
\Lambda\approx\dfrac{c}{2}\sqrt{\dfrac{(a\sqrt{1-e^2})^3}{3 G M}}\approx \sqrt{\dfrac{(a \sqrt{1-e^2})^3}{6R_S}},
\label{lambda_delta1}
\end{equation}
where $a$ is a semi-major axis and $e$ eccentricity for a selected orbit.
Therefore, after observations of bright stars for several decades an upper bound for a graviton mass could reach
around $5 \times 10^{-23}$~eV.

\subsection{Tidal charge constraints}

Reissner -- Nordstr\"om -- de-Sitter black hole could arise in theories with an extra dimension \cite{Dadhich_01}
 or in Horndeski scalar- tensor theories \cite{Babichev_17}.
The line element of the spherically symmetric Reissner -- Nordstr\"om -- de-Sitter  metric is
\beq
d s^2 = - f(r) d t^2 + f(r)^{-1} d r^2 + r^2 d\theta^2 + r^2 \sin^2\theta d\phi^2, \label{metric_RN}
\eeq
where  function $f(r)$ is defined as
\beq
f(r) = 1 - \dfrac{2M}{r}  + \dfrac{Q^2}{r^2} - \dfrac{1}{3} \Lambda r^2, 
\label{function_f}
\eeq
$M$ is a black hole mass, $Q$ is its charge and $\Lambda$ is cosmological constant.
In the case of a tidal charge \cite{Dadhich_01} or Horndeski scalar-tensor theories \cite{Babichev_17}, $Q^2$ could be negative.
For simplicity, we call $Q^2$ a tidal charge.
In paper \cite{Zakharov_EPJC_18} it was shown that a total relativistic advance for metric (\ref{metric_RN}) in PPN(1) approximation
could be expressed in an analytical form
\begin{equation}
  \Delta \theta (total) =   \frac{6\pi M}{L}                 -\frac{\pi Q^2}{ML}         +   \frac{\pi \Lambda a^3 \sqrt{1-e^2}}{M}.
\label{Delta_dS_2}
\end{equation}
As one can see  apocenter shift dependences on eccentricity and semi-major axis
are the same for Schwarzschild and Reissner -- Nordstr\"om cases while corresponding factors (${6\pi M}$  and    $-\dfrac{\pi Q^2}{M}$) are different, therefore,
it is very hard to distinguish a presence of a tidal charge and black hole mass evaluation uncertainties since black hole mass uncertainty could mimic
a existence of a tidal charge.
For $Q^2 > 0$, there is an apocenter shift in the opposite direction in respect to GR advance, while for $Q^2 > 0$ an apocenter shift  direction coincides with relativistic advance direction.
If we apply these relations for the supermassive black hole at the Galactic Center, in paper \cite{Zakharov_EPJC_18} bounds in $Q^2$ and $\Lambda$ are presented for current and future accuracies for Keck and Thirty Meter telescopes
which were discussed in \cite{Hees_PRL_17}.
Similarly to papers \cite{Zakharov_EPJC_18,Zakharov_Quarks_18} if we use   uncertainty  $\sigma_{\rm GRAVITY} = 0.030$~mas for the GRAVITY facilities as it was found in  \cite{Gravity_18} ($\delta_{\rm GRAVITY}=2\sigma_{\rm GRAVITY}$)
or  $\Delta \theta (GR)_{S2}= 13.84 \delta_{\rm GRAVITY} $ for S2 star and assuming again that GR predictions
about orbital precession of S2 star will be confirmed  with  $\delta_{\rm GRAVITY}$ accuracy (or $ \left|\dfrac{\pi Q^2}{ML}\right|  \lesssim \delta_{\rm GRAVITY}$), one could conclude that
\begin{equation}
   |Q^2| \lesssim 0.432 M^2,
\label{RN_GRAVITY}
\end{equation}
or based on results of  forthcoming observations with GRAVITY facilities one could expect to reduce significantly a possible range of $Q^2$ parameter in comparison with
a possible hypothetical range of $Q^2$ parameter in comparison with current and future Keck data.


\section{Conclusions}
As it was noted earlier,
precise observations of bright stars near the Galactic Center is very efficient tool  to check alternative theories of gravity and
to investigate a presence of an extended mass distribution near the Galactic Center. In our simple approach we compared
theoretical estimate for pericenter (apocenter) shifts with their estimates obtained from observations.
Using this approach we discussed an opportunities to evaluate parameters of supermassive black hole, stellar cluster and dark matter cloud near
the Galactic Center or evaluate parameters of alternative gravity model analyzing apocenter (pericenter) advance after at least one star revolution, however, in the future we will have a possibility to evaluate a static gravitational potential at the Galactic Center analyzing only  very small parts of stellar orbits
similarly to \cite{ONeil_18}, where it was shown even around 40\% of stellar phase coverage is enough for an orbit reconstruction.
However, if a contribution of time-dependent component of gravitational potential caused by stellar encounters or moving invisible mass
 concentrations
 is significant, then an orbit reconstruction problem may be more complicated.
One could obtain  the graviton mass constraint from an analysis of
S2 star trajectory and the bound is consistent and comparable with
the constraint presented recently by the LIGO collaboration, in particular, future observations of bright stars will give an opportunity to obtain
a graviton mass estimate which is slightly better than the best current graviton mass estimate found by the LIGO--Virgo collaboration.

If we assume that the supermassive black hole at the Galactic Center is described with Reissner -- Nordstr\"om -- de-Sitter metric, a tidal charge $Q^2$ (or corresponding parameter in  Horndeski gravity theories) could be constrained in an efficient way with monitoring the bright IR stars.

\subsection*{Acknowledgements}

The author thanks D. Borka, V. Borka Jovanovi\'c, F. De Paolis,  G.
Ingrosso, P. Jovanovi\'c, S. M. Kopeikin,  A. A. Nucita,  S.G. Rubin, B.
Vlahovic for useful discussions. A. F. Z. thanks also
NSF
(HRD-0833184) and NASA (NNX09AV07A) at NASA CADRE and NSF CREST
Centers (NCCU, Durham, NC, USA) for a partial support and the organizers of the Fourth International
Conference on Particle Physics and Astrophysics (ICPPA-2018) for their attention to our contribution.


\end{document}